\documentstyle[12pt]{article}

\input{epsf}

\newcommand{\xxxcA}{{\cal A}}
\newcommand{\wil}{W(\sigma)}

\newcommand{\half}{\frac{1}{2}}
\newcommand{\xxxcal}[1]{{\cal#1}}

\newcommand{\sect}[1]{\setcounter{equation}{0}\section{#1}}

\def\rf#1{(\ref{eq:#1})}
\def\lab#1{\label{eq:#1}}
\def\nonu{\nonumber}
\def\br{\begin{eqnarray}}
\def\er{\end{eqnarray}}
\def\be{\begin{equation}}
\def\ee{\end{equation}}

\def\lb{\lbrack}
\def\rb{\rbrack}

\def\({\left(}
\def\){\right)}

\newcommand{\ct}[1]{\cite{#1}}
\newcommand{\bi}[1]{\bibitem{#1}}

\relax

\newcommand{\sbr}[2]{\left\lbrack\,{#1}\, ,\,{#2}\,\right\rbrack}

\def\a{\alpha}

\def\G{\Gamma}

\def\l{\lambda}
\def\L{\Lambda}

\def\o{\over}

\def\pa{\partial}

\def\ra{\rightarrow}
\def\s{\sigma}
\def\S{\Sigma}

\def\tp0{\Theta_{+}^{(0)}}
\def\tm0{\Theta_{-}^{(0)}}

\def\u2{\mid u\mid^2}

\def\cg{{\cal G}}

\def\f#1#2#3 {f^{#1#2}_{#3}}

\def\win1{{\sf w_{1+\infty}}}

\def\Win1{{\sf W_{1+\infty}}}

\def\rlx{\relax\leavevmode}
\def\inbar{\vrule height1.5ex width.4pt depth0pt}
\def\IZ{\rlx\hbox{\sf Z\kern-.4em Z}}
\def\IR{\rlx\hbox{\rm I\kern-.18em R}}
\def\IC{\rlx\hbox{\,$\inbar\kern-.3em{\rm C}$}}
\def\IN{\rlx\hbox{\rm I\kern-.18em N}}
\def\IO{\rlx\hbox{\,$\inbar\kern-.3em{\rm O}$}}
\def\IP{\rlx\hbox{\rm I\kern-.18em P}}
\def\IQ{\rlx\hbox{\,$\inbar\kern-.3em{\rm Q}$}}
\def\IF{\rlx\hbox{\rm I\kern-.18em F}}
\def\IG{\rlx\hbox{\,$\inbar\kern-.3em{\rm G}$}}
\def\IH{\rlx\hbox{\rm I\kern-.18em H}}
\def\II{\rlx\hbox{\rm I\kern-.18em I}}
\def\IK{\rlx\hbox{\rm I\kern-.18em K}}
\def\IL{\rlx\hbox{\rm I\kern-.18em L}}
\def\one{\hbox{{1}\kern-.25em\hbox{l}}}
\def\0#1{\relax\ifmmode\mathaccent''7017{#1}%
B        \else\accent23#1\relax\fi}

\def\NPB#1#2#3{{\sl Nucl. Phys.} {\bf B#1} (#2) #3}

\def\PRD#1#2#3{{\sl Phys. Rev.} {\bf D#1} (#2) #3}

\def\PLB#1#2#3{{\sl Phys. Lett.} {\bf #1B} (#2) #3}
\def\JMP#1#2#3{{\sl J. Math. Phys.} {\bf #1} (#2) #3}

\def\TMP#1#2#3{{\sl Theor. Mat. Phys.} {\bf #1} (#2) #3}

\begin{document}

\begin{titlepage}
\vspace*{-1cm}
\noindent

\begin{center}
{\large\bf Integrable theories in any dimension: a perspective}\footnote
{Invited talk at the
 Meeting {\it Trends in Theoretical Physics II} at Buenos Aires, December 1998.}
\end{center}

\begin{center}
Orlando Alvarez$^{a}$, 
L.A. Ferreira$^{b}$
 and
J. S\'anchez Guill\'en$^{c}$

\vspace{.5 cm}
\small
\par \vskip .1in \noindent
$^{a}$ Department of Physics\\            
University of Miami\\              
P.O. Box 248046\\
Coral Gables, FL 33124, USA

\par \vskip .1in \noindent
$^{b}$Instituto de F\'\i sica Te\'orica - IFT/UNESP\\
Rua Pamplona 145\\
01405-900  S\~ao Paulo-SP, BRAZIL

\par \vskip .1in \noindent
$^{c}$Departamento de F\'\i sica de Part\'\i culas,\\
Facultad de F\'\i sica\\
Universidad de Santiago\\
E-15706 Santiago de Compostela, SPAIN
\normalsize
\end{center}
\vspace{.5in}
\begin{abstract}
We review the developments of a recently proposed approach to study integrable
theories in any dimension. The basic idea consists in generalizing the 
zero curvature representation for two-dimensional integrable models to 
space-times of  dimension $d+1$ by the introduction of a $d$-form connection. 
The method has been used to study several theories of physical interest,
like self-dual Yang-Mills theories, Bogomolny equations, non-linear sigma
models and Skyrme-type models.  The local version of the generalized zero
curvature involves a Lie algebra and a representation of it, leading to a
number of conservation laws equal to the dimension of that
representation. We discuss  the conditions a given theory has
to satisfy in order for its associated zero curvature to admit an infinite
dimensional (reducible) representation.  We also present the  theory in the
more abstract setting of 
the space of loops, which gives a deeper understanding  and a more simple 
formulation of integrability in any dimension. 
\end{abstract}
\end{titlepage}

\sect{Introduction}
\label{sec:intro}

This paper addresses the long standing problem of the generalization 
of the zero curvature in classical integrability, from two dimensions to
theories defined on space-times of any dimension. It consists of a review of
the ideas proposed in \ct{noso} to implement that generalization, and also of
the subsequent developments. 
 
Integrability and, in a more general sense, construction of solutions and 
constants of motion, are important issues, given 
the relevance of nonperturbative aspects of field theories in High Energy
Physics , Statistical Mechanics, Solid State Physics and Gravity.

In $2d$, an impressive understanding of integrable theories has been achieved,
which can be essentially encoded in  
the zero curvature formulation.  In fact, the equations of motion of such
theories, including relativistic invariant ones, can be naturally expressed in
terms of a flat  
connection as $ [\partial_\nu - A_\nu , \partial_\mu - A_\mu ]=0$.  
The simplest (abelian) example is that of a free scalar field $\phi$,  with  
$A_\mu=-  \epsilon_{\mu \nu } \partial^\mu\phi$.  The equations of motion of
the Sine-Gordon models and its 
generalizations, the so called Affine Toda systems, are also expressed as zero
curvature with $A_{\mu}$ taking values in an affine 
 algebra \ct{ot}.   
One of the remarkable facts is that the  algebraic
structure  involved in the zero curvature does not describe, in general, any
explicit symmetry of the original equations of motion.  In the case of
theories presenting soliton solutions, a great
unification is  achieved in the framework of affine  algebras, where 
there exists algebraic equivalent methods, like dressing of vacuum solutions 
or solitonic specialization giving the interesting solitonic 
solutions  in a systematic and simple way \ct{nos}.
 
The zero curvature formulation also provides  a way of constructing 
conserved charges. This is easily seen choosing a 
finite two dimensional space-time  with periodic boundary conditions on the
time component of the connection.  By Stokes calculus, any power $N$ of the 
 $ Tr (P\exp (\int_{-L} ^LÊ A_x (x,t)dx))^N$ is conserved in the 
time evolution (the path ordering $P$ is for the non-abelian general 
case).  Despite the great advances in  $2d$ classical integrability, and also
on some relations between integrability in {\it parameter space} 
and solutions of supersymmetric gauge 
theories in $4d$ \ct{sw}, very little was done in  generalizing this 
beautiful picture to any dimension in {\it space time}, and understand 
{\it directly} those solutions in terms of generalized soliton dynamics, 
most interesting by itself.  But expressing relativistic invariant field 
equations, known and possibly new ones, in such geometrical way which 
yields their integrals of motion, cannot be simple in higher 
dimensions.  First of all, relativistic invariance requires higher 
rank tensors and connections, in principle with complicated  gauge 
transformations.  Besides, in $2d$ the Lorentz 
transformations are rather trivial $ x_\pm \rightarrow \lambda ^{\pm 
1}x_\pm$ and they can be imitated by the grading operator in the 
algebra, just making the forward (backward) light-cone to be positive 
(negative), incorporating also holomorphicity directly.  

The main difficulties in extending of the integrability concepts to higher
dimensions are associated to non-locality issues that rise when dealing with
higher rank connections. Those problems can be circunvented by the
introduction of auxiliary connections that allow for parallel transport. In
addtion, once the 
invariant equations have been formulated, to get their integrals of 
motion one needs a generalization of Stokes calculus, which is 
directly related to the first problem of the transformation of the 
generalized connections.  
In fact Stokes formulae are problematic for the relevant non-abelian 
case even in $2d$ \ct{stokes}. Therefore, the starting point of \ct{noso} was 
to obtain a  simple local expression for the non-abelian Stokes theorem, as we
review in section \ref{sect:curv}. 

Our approach to generalize the zero curvature is based on the  observation
that the conservation laws can be expressed as the condition for the
invariance of some quantity under the deformation of hypersurfaces. We are in
fact constructing higher dimensional non-abelian Gauss-type laws. The main
difficult is related to the implementation of the calculus of the variation of
the hypersurfaces and the requirement that at the end one wants {\em local}
conservation laws. The key observation for implementing the calculus was based
on the formula satisfied by the Wilson loop $W_c$, calculated on a closed
countour $C$, i.e. 
\be
{d W_c\o{dt}} = W_c T(F, 2 \pi ,t) 
\lab{veq1} 
\ee 
where $t$ parametrizes the variations of the countour that keep a given
fixed point $x_0$ fixed, and   
\be 
T(F, 2 \pi ,t) \equiv \int_0^{2\pi} d\s W^{-1} 
F_{\mu\nu} W {d x^{\mu}\o{d\s}} {d x^{\nu}\o{dt}},  
\ee 
and 
$ F_{\mu\nu}$ is the curvature of the connection $A_{\mu}$ defining the Wilson
line $W$, i.e. 
\be
{d W\o{d\s}} = - A_{\mu} {d x^{\mu}\o{d\s}} W 
\ee
and the  integral in $T$ is along the 
curve parameter $\sigma $.    Eq.\rf{veq1}, which is easily obtained from
Interaction  
Picture method in \ct{noso}, expresses loop independence when the 
connection is flat and it is essentially non-local. This simple formulation
gives the idea for the zero curvature generalization to $3d$: introduce a 
functional of the fields $V$ defined by its variation under deformations of 
the loops through the equation 
\be
{d V\o{dt}} = V T(B, 2 \pi ,t) 
\ee
where $B_{\mu\nu}$ is a general antisymmetric tensor 
functional and 
\be
T(B, 2 \pi ,t) \equiv \int_0^{2\pi} d\s W^{-1} 
B_{\mu\nu} W {d x^{\mu}\o{d\s}} {d x^{\nu}\o{dt}}, 
\ee
One looks then for the conditions for $V$, now a surface integral, to be 
independent of the surface scanned by the different loops.  Notice 
that the integrand in $ T(B, 2 \pi ,t)$ is a local function.  The form 
of $T$ will be essential to guarantee Lorentz and gauge invariant time 
evolution.  Generalizing directly the previous 2d Interaction picture 
computation varying now with respect to a new orthogonal parameter 
$\tau$, we obtain a rather complicated expression, 
albeit with a transparent geometrical and physical meaning.  It is a 
generalized non-abelian Gauss law and, as shown in detail in section 
\ref{sub:3dzc}, it corresponds to parallel transport in the space of loops.
This geometrical theory  tell us how 
to  proceed in any dimension and makes very natural the translation 
of the surface independence to different {\it local} equations of 
motion and/or integrability conditions, which are deduced directly in 
section \ref{sect:local}.  A natural condition is the flatness of the first 
connection $F_A =0$, which guarantees independence of how the surface is 
scanned by loops,  and then a sufficient condition for the vanishing of the
non-local curvature associated to $V$  is the constant 
covariance of the tensor connection $D_A B=0$.  This nesting of constant 
covariance and vanishing of the commutator of two $d -forms$ and 
flatness of the lower $d-1$, $d-2$...  curvatures (or possibly some of 
them) is generic and it is worked out in detail also for the $4d$ (see
\ct{noso} for details). 

The Lorentz covariant and gauge invariant formalism is very general and one 
has lots of algebraic and topological structures to explore which 
support those local formulations and which will yield new formulations 
of known theories or new ones, with a systematic method to get their 
integrals of motion and solutions.  In $3d$, the most simple case is 
choosing $B_{\mu\nu}(0)$ on an abelian subalgebra.  Notice that 
$B_{\mu\nu}$ can then be uniquely defined at any point $x$ by parallel
transport under $A_{\mu}$.  The  
integrability conditions in this case correspond to the BF theory, 
Chern-Simons and other ``topological'' theories in 3d.  That the 
construction works for these topological cases is reassuring and one 
can define now naturally new observables like invariants for link 
crossings or study topological defects, but the importance of 
conserved quantities is, of course, with time evolution.  On the other 
hand very little is clearly understood about non-perturbative 
solutions and classical integrability for the situation when there is 
full relativistic dynamics in dimension higher than 2.  So we concentrated in
\ct{noso} on the simplest 
genuine $2+1$ theory, the $O(3)$ nonlinear model, although most of the 
approach can be used for a general sigma model as shown later in \ct{fl,mad}.  
In  the four dimensional case, it was shown in \ct{noso} how to express
self-dual Yang-Mills and BPS theory in such 
zero curvature formulation of local integrability conditions for 
volume independence.  We obtain also a reduction of the 
equations which exhibit directly their conservation character.

One of the interesting aspects of \ct{noso} is that many theories 
presenting the local  zero curvature are not integrable in the sense of
possessing an  
infinite number of conservation laws. However, some of those theories 
contain integrable submodels that do present an infinite number of
conserved currents. We discuss in section \ref{sect:intsub} the conditions a
given theory has to satisfy to present an infinity number of conserved local
currents. The main point is that its zero curvature representation has to
involve infinite dimensional representations of a Lie algebra, or equivalently
as we explain, and infinite dimensional non-semisimple Lie algebra of the
Poincar\'e type.

\sect{Geometrical approach to integrability}
\label{sect:curv}

The zero curvature condition in two dimensional spacetime, know as the
Zakharov-Shabat equation \ct{zs} is given by
\begin{equation}
F_{\mu\nu} \equiv [\partial_{\mu} + A_{\mu} , \partial_{\nu} + A_{\nu} ]=0 \; 
\qquad \qquad \mu \, ,\nu =0,1 
\lab{2dint}
\end{equation}
One of the consequences of it is that it leads to conservation
laws. In order to see that, consider a quantity $W$  defined by paralell
transport with a connection $A_{\mu}$, 
\be
{d W\o{d\s}} + A_{\mu} {d x^{\mu}\o{d\s}} W = 0
\lab{weqintro}
\ee
Consider now a closed curve $\G$, and let $\S$ be a
two dimensional surface having $\G$ as its boundary. 
The nonabelian Stokes theorem states that the quantity $W$ can be determined
either by integrating \rf{weqintro} on $\G$ or on $\S$. More precisely one has 
\begin{equation}
W\( \G \) = P\exp\( \int_{\G} d\s A_{\mu} {dx^{\mu}\o d\s} \) = 
{\cal P} \exp \( \int_{\S} d\tau\,  d\s W^{-1} F_{\mu\nu} W 
{d x^{\mu}\o{d\s}}{d x^{\nu}\o{d\tau}}\)
\lab{stokestheor}
\end{equation}
where $P$ and ${\cal P}$ mean path and surface ordering respectively (see
\ct{noso} for details).

Therefore, if the connection $A_{\mu}$ is flat, i.e. \rf{2dint} holds true,
then $W$ is equal to unity for closed loops. Then, conserved quantities are
constructed as follows.  First we consider  
the case where spacetime is a cylinder.  At a fixed time $t_0$ 
consider a loop $\gamma_0$ beginning and ending at $x_0$.  At a later 
fixed time $t_1$ consider a loop $\gamma_1$ also beginning and ending 
at $x_0$.  Let $\gamma_{01}$ be a path connecting $(t_0,x_0)$ with 
$(t_1,x_0)$.  The flat connection allows us to integrate the parallel 
transport equation \rf{weqintro} along two different paths obtaining 
$W(\gamma_0) = W(\gamma_{01})^{-1} W(\gamma_1) W(\gamma_{01})$.  We 
first observe that $W(\gamma_0)$ transforms under a gauge 
transformation $g(x)$ as $W(\gamma_0) \to g(x_0) W(\gamma_0) 
g(x_0)^{-1}$.  The conserved quantity should be gauge invariant.  If $\chi$ 
is a character for the group $G$ we have that $\chi(W(\gamma_0))$ will be 
gauge invariant.  Also $\chi(W(\gamma_0))= \chi(W(\gamma_1))$.  Thus 
we can construct a conserved gauge invariant quantity
\begin{equation}
	\chi(W(\gamma_0)) 
\lab{conserv2d}
\end{equation}
for every independent character of the group. These are the constants 
of motion in the zero curvature construction. Note that the data needed to 
compute $\chi(W(\gamma_0))$ is all determined at time $t_0$.

In the case where the spacetime is two dimensional Minkowski space one 
has to impose physically sensible boundary conditions at spatial 
infinity.  Note that $\;$ $P \exp \left(\int_{-\infty}^\infty A_x dx\right)$ 
is not gauge invariant if one allows nontrivial gauge transformations 
at infinity.  In setting up the problem one has to choose the correct 
physical boundary conditions which may for example require that $A_0$ 
vanishes at infinity.  Depending on the details one can construct a 
conserved quantity by a slight modification of the construction above.

The conserved quantities \rf{conserv2d} are nonlocal because in general the 
connection $A_{\mu}$ lies in a nonabelian algebra. However, in cases like the 
affine Toda models \ct{ot}, it 
is possible to get local conservation laws by gauge transforming $A_{\mu}$
into an abelian subalgebra.

The basic idea in \ct{noso} to bring such concepts to higher dimensions, is 
to introduce quantities integrated over hypersurfaces and to find the 
conditions for them to be independent of deformations of the hypersurfaces 
which keep their boundaries fixed. Such an approach will certainly lead to 
conservation laws in a manner very similar to the two dimensional case. 
However, the main problem of that it is how to introduce non-linear zero 
curvatures keeping things as local as possible. The way out is to introduce 
auxiliary connections to allow for parallel transport. The number of 
possibilities  of implementing those ideas increase with the dimensionality 
of space-time. However, the simplest scenario is that where, in a space-time 
of dimension $d+1$, one introduces a rank $d$ antisymmetric tensor 
$B_{\mu_1\mu_2 \ldots \mu_d}$ and a vector $A_{\mu}$.   
The idea can perhaps be best stated using a formulation in ``loop space''. 
On a $d+1$ dimensional space-time $M$ one considers the space 
$\Omega^{d-1}(M,x_0)$ of $d-1$ dimensional closed hypersurfaces based at a 
fixed point $x_0 \in M$. One then introduces on such ``higher loop space'' 
a  $1$-form $\xxxcal{A}$ which is basically the quantity  
$W^{-1}B_{\mu_1\mu_2 \ldots \mu_d}W$ integrated over the closed hypersurfaces 
(see \ct{noso} for details). The quantity $W$ is defined in terms of the 
vector $A_{\mu}$ through \rf{weqintro}. However, for $W$ to be independent of 
the way one integrates it from $x_0$ to a given point on the hypersurface, 
one has to assume that $A_{\mu}$ is flat, i.e.
\be
F_{\mu\nu} = [ D_{\mu} , D_{\nu} ] = \partial_{\mu}\, A_{\nu} - 
\partial_{\nu}\, A_{\mu} + [A_{\mu} , A_{\nu} ]=0   \; ; 
\qquad \mu , \nu = 0,1,2 \ldots d 
\lab{stzc}
\ee
with
\be
D_{\mu} \cdot \equiv \partial_{\mu}\, \cdot  + [A_{\mu} \, , \, \cdot \, ]
\lab{covder}
\ee

Roughly speaking a $d$ dimensional 
closed hypersurface in $M$, based at $x_0$, corresponds to a (one 
dimensional) loop in $\Omega^{d-1}(M,x_0)$. Therefore, the condition to have 
things independent of deformation of hypersurfaces translates in such 
``higher loop space'' to the zero curvature condition for $\xxxcal{A}$, namely 
\be
\xxxcal{F} = 
\delta\xxxcal{A} + \xxxcal{A}\wedge\xxxcal{A} = 0
\lab{loopzc}
\ee
The relation \rf{loopzc} (together with \rf{stzc}) is the generalization of 
the zero curvature \rf{2dint} to higher dimensions proposed in \ct{noso}.

The idea of constructing conserved quantities using \rf{loopzc} is the same as
the two dimensional case decribed above. Indeed, the condition \rf{loopzc}
implies that the path ordered exponential of $\xxxcal{A}$ along a closed loop
in the higher loop space $\Omega^{d-1}(M,x_0)$ is independent of the
loop. Following arguments similar to those below \rf{stokestheor}, one sees
that the 
conserved quantities are path ordered exponentials of $\xxxcal{A}$ along
 closed loop at a given fixed time. Since loops in $\Omega^{d-1}(M,x_0)$
correspond to $d$-dimesnional closed surfaces in the $d+1$-dimensional
spcaetime $M$, one gets that such quantities are indeed integrated over the
physical space. Although the analogy with the two dimensional case looks
straitghforward, the implementation is quite involved due to the ordering of
the integration, the respect to gauge invariance and boundary conditions. We
refer to \ct{noso} for those very important explanations. 

In order to make the formulas more explicit we give below the discussion of
the $2+1$ dimensional case.

\subsection{The zero curvature in three dimensions}
\label{sub:3dzc}

Assume we have a principal bundle $P\to M$ with connection.  For a 
fixed point $x_0\in M$ let $\Omega(M,x_0)$ be the space of all loops 
based at $x_0$:
\be
	\Omega(M,x_0) = \{ \gamma: S^1 \to M \mid \gamma(0)=x_0\}\;.
\lab{loopspacedef}
\ee
we now want to construct a principal $G$-bundle over $\Omega(M,x_0)$ 
with connection.  Note that the structure group of the bundle is a 
finite dimensional group not a loop group.  It will be the trivial 
bundle $\xxxcal{P}= \Omega(M,x_0)\times G$.  Conceptually the bundle 
is constructed as follows.  Over $x_0 \in M$ the bundle $P\to M$ has 
fiber $P_{x_0}$ which is isomorphic to $G$.  All loops in 
$\Omega(M,x_0)$ have $x_0\in M$ as a starting point so we can consider 
them having $P_{x_0}$ in common.  This is the common fiber in the 
cartesian product $\Omega(M,x_0)\times G$.  Mathematically we have a 
natural map $\pi:\Omega(M,x_0) \to M$ given by $\pi(\gamma)=x_0$.  The 
bundle $\xxxcal{P}$ is just the pullback bundle $\pi^* P$, see 
\ct{nepo}.  Since 
$\xxxcal{P} \to \Omega(M,x_0)$ is a trivial bundle we can put the 
trivial connection on it.  There is a more interesting connection one 
can put on it which exploits the connection on the bundle $P\to M$.  
Consider a Lie algebra valued $2$-form $B$ on $M$ such that under the 
transition function $\phi$ we have $B \to \phi^{-1}B\phi$.  Let 
$W(\sigma)$ be the parallel transport operator from the point 
$x(0)=x_0$ to the point $x(\sigma)$ along the loop $\gamma$.  We can 
assign the Lie algebra valued $1$-form
\begin{equation}
	\xxxcal{A}[x(\sigma)] = \int_0^{2\pi} d\sigma 
	\; W(\sigma)^{-1}B_{\mu\nu}(x(\sigma))W(\sigma)
	\frac{dx^\mu}{d\sigma}
	\delta x^\nu(\sigma)
	\label{calA}
\end{equation}
The transformation laws of the above are determined by $\phi(x_0)$ 
which is clearly associated with the common fiber $P_{x_0}$. Thus we can 
define a connection on $\xxxcal{P}$ by
$$
	\varpi = -dg g^{-1} + g\left( \int_0^{2\pi} d\sigma 
	\; W(\sigma)^{-1}B_{\mu\nu}(x(\sigma))W(\sigma)
	\frac{dx^\mu}{d\sigma}
	\delta x^\nu(\sigma)
		 \right) g^{-1} \;.
$$
Thus we can treat $\xxxcal{A}$ as the connection in a certain 
trivialization.  The curvature is given by $\xxxcal{F} = 
\delta\xxxcal{A} + \xxxcal{A}\wedge\xxxcal{A}$. 

What do we mean when we say that we want to have parallel transport 
independent of path in $\Omega(M,x_0)$? Look at the space 
$\Omega(M,x_0)$ and consider a curve $\Gamma$ in $\Omega(M,x_0)$ 
parametrized by $\tau$ such that $\Gamma(0)$ is the ``constant curve'' 
$x_0$.  Note that for fixed $\tau$, $\Gamma(\tau)$ is a curve 
$x_\tau(\sigma)$ for $\sigma\in [0,2\pi]$ in $M$.  Thus it is convenient 
to ``write'' $\Gamma$ as $x(\sigma,\tau)$.  The statement that parallel 
transport be independent of the choice of curve $\Gamma \in 
\Omega(M,x_0)$ with fixed starting and ending points is the statement 
that the curvature vanish.
If one wants the parallel transport between points in $\Omega(M,x_0)$ 
to be independent of path then parallel transport should be path 
independent in $M$.  The reason for this is that a loop in 
$\Omega(M,x_0)$ beginning at the trivial loop may be viewed as a map 
from the square $[0,2\pi]^2$ to $M$ such that $\partial [0,2\pi]^2$ 
gets mapped to $x_0$.  
To get the same result for two different 
sets of ``constant'' $\tau$ curves associated with the same closed 
$2$-submanifold in $M$, one needs ordinary parallel 
transport to be path independent, i.e.  $F=0$.

In order to perform  the curvature computation we need 
the standard result:
\begin{eqnarray*}
	 W(\sigma)^{-1}\delta W(\sigma)& = & 
	 - W(\sigma)^{-1} A_\mu(x(\sigma)) 
	W(\sigma) \delta x^\mu(\sigma)  \\
	 & + & \int_0^\sigma d\sigma'\;
	W(\sigma')^{-1} F_{\mu\nu}(x(\sigma'))W(\sigma') \frac{dx^\mu}{d\sigma'}
	\delta x^\nu(\sigma')\;.
\end{eqnarray*}
We also need definition~(\ref{calA}). 
Let $\delta$ be the exterior derivative on the space 
$\Omega(M,x_0)$ and thus $\delta^2=0$ and 
$$
	\delta x^\mu(\sigma)\wedge \delta x^\nu(\sigma') 
	=- \delta x^\nu(\sigma')\wedge\delta x^\mu(\sigma)\;.
$$

Computing the curvature $\xxxcal{F} = \delta\xxxcA + \xxxcA\wedge\xxxcA$ 
is  tedious and we refer to \ct{noso} for the details. The result is 
\begin{eqnarray*}
	\xxxcal{F}
	 & = & - \half\int_0^{2\pi} d\sigma \;
	 \wil^{-1} 
	 \left[ D_\lambda B_{\mu\nu}
	 + D_\mu B_{\nu\lambda}
	 + D_\nu B_{\lambda\mu}\right](x(\sigma)) 
	 \wil \\
	 & &  \mathstrut\hspace{.5in} \times
	 \frac{dx^\lambda}{d\sigma}
	 \delta x^\mu(\sigma)\wedge \delta x^\nu(\sigma) \\
	& - & \int_0^{2\pi} d\sigma \int_0^\sigma d\sigma' \;
	 \left[ F^W_{\kappa\mu}(x(\sigma')) ,
	 B^W_{\lambda\nu}(x(\sigma)) \right] 
	\frac{dx^\kappa}{d\sigma'} \frac{dx^\lambda}{d\sigma}
	\delta x^\mu(\sigma') \wedge\delta x^\nu(\sigma) \\
	& + &
	\half \int_0^{2\pi} d\sigma \int_0^{2\pi} d\sigma'
	\left[ B^W_{\kappa\mu}(x(\sigma')), B^W_{\lambda\nu}(x(\sigma))
	\right]
	\frac{dx^\kappa}{d\sigma'}\frac{dx^\lambda}{d\sigma}
	\delta x^\mu(\sigma') \wedge \delta x^\nu(\sigma)\;.
\lab{curvatural}
\end{eqnarray*}
The vanishing of $\xxxcal{F}$, i.e. relation \rf{loopzc}, 
is our generalization of the zero curvature to higher dimensions. It implies
hypersurface indepence and conservation laws as we explained above.

\sect{Local integrability conditions}
\label{sect:local}

The condition \rf{loopzc} that the loop space curvature  should vanish is local
in $\Omega^{d-1}(M,x_0)$, but it is highly  
non-local in the spacetime $M$.  We now discuss some sufficient local
conditions for the vanishing of \rf{curvatural}. 

Let $\cg$ be a Lie algebra and $R$ be a representation of it.  
We introduce the nonsemisimple Lie algebra $\cg_R$ as
\br
\lb T_a \, , \, T_b \rb &=& f_{ab}^c T_c \nonu\\
\lb T_a \, , \, P_i \rb &=& P_j R_{ji}\( T_a\) \nonu\\
\lb P_i \, , \, P_j \rb &=& 0
\lab{rt}
\er
where $T_a$ constitute a basis of $\cg$ and $P_i$ a basis for the abelian
ideal $P$ (representation space). The fact that $R$ is a matrix
representation, i.e.
\begin{equation}
\lb R\( T_a \) \, , \, R\( T_b\)  \rb =  R\( \lb T_a \, , \, T_b \rb \)
\lab{rep}
\end{equation}
follows from the Jacobi identities.

We take the connection $A_{\mu}$ to be in $\cg$ and the rank $d$ 
antisymmetric tensor $B_{\mu_1\mu_2 \ldots \mu_d}$ to be in $P$, i.e.
\begin{equation}
A_{\mu} = A_{\mu}^a T_a \; , \qquad B_{\mu_1\mu_2 \ldots \mu_d} = 
B_{\mu_1\mu_2 \ldots \mu_d}^i P_i 
\lab{abalg}
\end{equation} 

Then a set of sufficient {\em local} conditions for the vanishing of the 
curvature $\xxxcal{F}$ in \rf{loopzc} is given by 
\be
D_{\mu} {\tilde B}^{\mu} = 0 \; ; \qquad F_{\mu\nu} = 0 
\lab{localzc}
\ee
where we have introduced  the dual of $B_{\mu_1\mu_2 \ldots \mu_d}$ as 
\be
{\tilde B}^{\mu} \equiv {1\o d!} \, 
\varepsilon^{\mu \mu_1\mu_2 \ldots \mu_{d}} \, B_{\mu_1\mu_2 \ldots \mu_d}
\lab{dual}
\ee

Indeed, in the $2+1$ dimensional case the conditions  \rf{rt} and \rf{abalg}
imply that 
\be
W^{-1} B_{\mu\nu} W \in P
\ee
and therefore the commutator in the last term of \rf{curvatural} vanishes,
since $P$ is abelian. The first condition in \rf{localzc} in the $2+1$
dimensional case reads 
\begin{equation}
D_{\l} B_{\mu\nu} + D_{\mu} B_{\nu\l} +  D_{\nu} B_{\l\mu} = 0 
\lab{covconserb}
\end{equation}
Therefore, together with $F_{\mu\nu} = 0$, it implies that the remaining terms
of \rf{curvatural} vanish too.  

The relations \rf{localzc}  are the {\em local} integrability conditions which
we introduce for theories 
defined on  a spacetime of any dimension,
and which constitutes a generalization of the zero curvature condition
\rf{2dint} in two dimensions. They lead to local 
conservation laws. Indeed, since the connection $A_{\mu}$ is flat it can be 
written as 
\be
A_{\mu} = -\partial_{\mu} W \, W^{-1}
\lab{puregauge}
\ee
and consequently \rf{localzc} imply that the currents
\be
J_{\mu} \equiv  W^{-1}\, {\tilde B}_{\mu} \, W 
\lab{currents}
\ee
are conserved
\be
\partial_{\mu} \, J^{\mu} = 0
\lab{conserv}
\ee

The zero curvature conditions \rf{localzc} are invariant under the gauge 
transformations
\br
A_{\mu} &\ra & 
g \, A_{\mu} \, g^{-1} - \pa_{\mu} g \, g^{-1} \nonu\\
{\tilde B}_{\mu} &\ra &  
g \, {\tilde B}_{\mu} \, g^{-1} 
\lab{gauge}
\er
and 
\br
A_{\mu} &\ra & A_{\mu} \nonu\\
{\tilde B}_{\mu} &\ra & {\tilde B}_{\mu} + 
\varepsilon_{\mu\mu_1 \ldots \mu_d} D^{\mu_1} \a^{\mu_2 \ldots \mu_d} \equiv 
{\tilde B}_{\mu} + D^{\nu} {\tilde \a}_{\mu\nu}
\lab{newgauge}
\er
where we have introduced the dual ${\tilde \a}_{\mu\nu} \equiv 
\varepsilon_{\mu\nu\mu_2 \ldots \mu_d}\a^{\mu_2 \ldots \mu_d}$. 
In \rf{gauge} $g$ is an element of the group obtained by exponentiating the 
Lie algebra $\cg$. The transformations \rf{newgauge} are symmetries of 
\rf{localzc} as a consequence of the fact that the connection $A_{\mu}$ is 
flat, i.e. $\lb D_{\mu} \, , \, D_{\nu}\rb = 0$. In addition, the parameters 
$\a^{\mu_1 \ldots \mu_{d-1}}$ take values in the abelian ideal $P$. 

The currents \rf{currents} are invariant under the transformations 
\rf{gauge}, but under \rf{newgauge} they transform as 
\be
J_{\mu} \ra J_{\mu} + 
\varepsilon_{\mu\mu_1 \ldots \mu_d} \partial^{\mu_1}\( W^{-1} \, 
\a^{\mu_2 \ldots \mu_d}\, W \) = J_{\mu} + 
\partial^{\nu}\( W^{-1} \,{\tilde \a}_{\mu\nu} \, W \) 
\lab{curtransf}
\ee

The transformations \rf{gauge} and \rf{newgauge} do not commute and their 
algebra is isomorphic to the non-semisimple algebra $\cg_R$ introduced 
in \rf{rt}. The nontrivial gauge transformations allow in principle the
dressing of vacuum solutions to obtain general ones, as discussed in \ct{noso}.

\sect{Integrable submodels}
\label{sect:intsub} 

The number of conserved currents one gets from \rf{currents} is equal to the
dimension of the representation of $\cg$, defined by the generators $P_i$ of
$\cg_R$. Consequently, the notion of integrability in such approach is related
to infinite dimensional representations, or equivalently to
infinite dimensional non-semisimple Lie algebras of the type \rf{rt}. That is
similar to the  two dimensional case where the appearence of infinte number of
charges is also linked to infinite dimensional Lie algebras. However, those 
 are in general of the affine  type, and it is now quite well understood
the role they have  in soliton
theory and exact methods of construction of solutions. In our approach, the
role of algebraic structures like \rf{rt} is not fully understood yet,
but we believe it must have profound consequences in the study of higher
dimensional integrable theories. 

One point that became clear recently is that of integrable submodels. It is
known that 
conditions like self-duality in Yang-Mills theories and the Bogomolny
equations in gauge theories with symmetry spontaneously broken by a Higgs in
the adjoint representation, play a crucial role in the construction of
submodels which possess properties of solvability not present in the full
theory. Those conditions lead to a saturation of a bound on the Euclidean
action in the case of self-dual Yang-Mills and on the energy in the case of
Bogomolny equations. 

In our approach a similar thing happens, however involving  quite
different structures. It does lead to conditions for integrable submodels, but
apparently not to saturation of bounds. In the examples studied so far what
one gets is the following. Suppose that using the gauge symmetries \rf{gauge}
and \rf{newgauge} one can find a gauge where the connection $A_{\mu}$ can be
split as \ct{fl} 
\be
A_{\mu} = A_{\mu}^{S} + A_{\mu}^{K}
\ee
where $A_{\mu}^{S/K}$ are the components of $A_{\mu}$ in the decomposition 
\be
\cg = S + K
\ee
with $K$ being a subalgebra of $\cg$, and $S$ its complement in $\cg$. Suppose
in addition that 
\be
\sbr{ A_{\mu}^{S}}{{\tilde B}^{\mu}} = 0
\ee
Then the first zero curvature  \rf{localzc} becomes
\be
\pa^{\mu}{\tilde B}_{\mu} + \sbr{{A^{\mu}}^{K}}{{\tilde B}_{\mu}} = 
\( \pa^{\mu}{\tilde B}_{\mu ,i} + {A^{\mu}_a}^{K}{\tilde B}_{\mu ,j}
R\( K_a\)_{ij} \) P_i 
\lab{zcmodel}
\ee
where ${A^{\mu}}^{K} = {A^{\mu}_a}^{K} K_a$, with $K_a$ being the generators of
$K$, and 
$R\( K_a\)_{ij}$ being the matrix representation of $K$ defined by the
$P_i$'s, i.e.
\be
\sbr{K_a}{P_i} = P_j R\( K_a\)_{ji}
\ee

Therefore, the zero curvature condition is only determined by the
representation of $K$ defined by the subspace $P$. Such a representation is in
general reducible, and we shall denote the branching as 
\be
R = \sum_l R_l^K
\ee
with $R_l^K$ being irreducible representations of $K$.  
Suppose now that a given representation $R^{\l}$ of $\cg$ presents a 
 branching rule 
\be
R^{\l} = \sum_l R_l^K + {\rm something}
\lab{condrep}
\ee
Then we can introduce an operator 
\be
{\tilde B}_{\mu}^{\l} \equiv {\tilde B}_{\mu ,i}\, P_i^{\l} 
\ee
where ${\tilde B}_{\mu ,i}$ are the same coefficients as in the expansion of
the original ${\tilde B}_{\mu}$ in terms of the basis $P_i$ of the
representation $R$ of $\cg$, i.e. ${\tilde B}_{\mu} = {\tilde B}_{\mu ,i}\,
P_i$. In addition, $P_i^{\l}$ are the generators of $\cg_{R^{\l}}$ associated
to the representation $R^{\l}$, 
corresponding to the subspace $\sum_l R_l^K$, and transforming exactly as the
$P_i$'s under the subalgebra $K$, i.e.
\be
\sbr{K_a}{P_i^{\l}} = P_j^{\l} R\( K_a\)_{ji}
\ee

Therefore, one gets that
\br
D^{\mu}{{\tilde B}_{\mu}}^{\l} &=& \pa^{\mu}{{\tilde B}_{\mu}^{\l}} 
+ \sbr{{A^{\mu}}}{{\tilde B}_{\mu}^{\l}} \nonu\\
&=& \( \pa^{\mu}{\tilde B}_{\mu ,i} + {A^{\mu}_a}^{K}{\tilde B}_{\mu ,j}
R\( K_a\)_{ij} \) P_i^{\l} \nonu\\
&+& \sbr{{A^{\mu}}^{S}}{{\tilde B}_{\mu}^{\l}}
\lab{fullzclambda}
\er

Notice that the first term after the last equality in \rf{fullzclambda}
corresponds to the zero curvature for the theory under consideration, namely
\rf{zcmodel}. Consequently, if one imposes the constraint 
\be
\sbr{{A^{\mu}}^{S}}{{\tilde B}_{\mu}^{\l}} = {A^{\mu}}^{S}_r {\tilde
B}_{\mu}^i \sbr{S_r}{P_i^{\l}} = 0
\lab{constaintgen}
\ee
with $S_r$ being the generators of the subspace $S$, one gets a submodel with
the conserved currents given by (see \rf{currents})  
\be
J_{\mu}^{\l} \equiv  W^{-1}\, {\tilde B}_{\mu}^{\l} \, W 
\lab{currentslambda}
\ee

Suppose now that there exists an infinite number of representations, like
$R^{\l}$, satisfying 
\rf{condrep} such that the conditions \rf{constaintgen} impose the same set
of constraints on the model. Then the submodel defined by the equations
\rf{zcmodel} and the constraints \rf{constaintgen} possesses an infinite
number of local conserved currents. Several examples fullfilling those
requirements were constructed in \ct{noso,mad,fl,afz}.

An important case where the above construction works is when the
representation of $\cg$ defined by the $P_i$'s, possesses at least one charge
zero singlet of the subalgebra $K$, i.e. there exists in the abelian
subalgebra $P$ a generator $P_{\L}$, such that
\ct{fl} 
\be
\sbr{K_a}{P_{\L}}=0
\ee

Then one can easily construct representations satisfying \rf{condrep} by
considering tensor products of the representation $R$, defined by the
$P_i$'s, with itself. The subsapces given by the tensor products of $R$ with
the singlet $P_{\L}$, transform like $\sum_l R_l^K$.  For instance, in the
case of  $R\otimes R$ one has that $P_{\L}\otimes R$ (equivalently $R\otimes
P_{\L}$) satisfies
\be
\sbr{1\otimes K_a + K_a \otimes 1}{P_{\L}\otimes P_i} = 
P_{\L}\otimes P_j \, \, \, R\( K_a\)_{ji}
\ee

For the case of $\(\otimes R\)^n$ any representation of the
form  
$\(\otimes P_{\Lambda}\)^l \otimes R\(\otimes
P_{\Lambda}\)^{n-l-1}$  is equivalent to
$R$ (viewed as  representations of the subalgebra $K$) . Therefore,  one
introduces the potentials  
\br
A_{\mu}^{(n)} &\equiv&  A_{\mu}^{\a} \sum_{l=0}^{n-1} 
\(\otimes 1\)^l \otimes T_{\a} \(\otimes 1\)^{n-l-1}\nonu\\
{\tilde B}_{\mu}^{(n)} &\equiv&   {\tilde B}_{\mu ,i } \, \sum_{l=0}^{n-1}
c_{n,l}\;  \(\otimes  P_{\Lambda}\)^l \otimes P_i 
\(\otimes  P_{\Lambda}\)^{n-l-1}
\lab{cosetpottensor}
\er
where we have denoted $A_{\mu} =
A_{\mu}^{\a} \, T_{\a}$, with $T_{\a}$ being the generators of $\cg$, and
where $c_{n,l}$ are constants.  We
introduce such constants because one can 
rescale the basis of each irreducible 
component of the re\-pre\-sen\-ta\-tions of $K$ independently, without
affecting the 
equations \rf{zcmodel}. Only the constraints, defining the submodel, are
affected by the constants $c_{n,l}$. 

Notice that the curvature $F^{(n)}_{\mu\nu}$ associated to the connection
$A_{\mu}^{(n)}$ vanishes as a consquence of the fact that $A_{\mu}$ is
flat. Therefore, the first condition in \rf{localzc}   
leads in this case, to the same equations as following equations as
\rf{zcmodel}, i.e. 
\be
\pa^{\mu}{\tilde B}_{\mu ,i} + {A^{\mu}_a}^{K}{\tilde B}_{\mu ,j}
R\( K_a\)_{ij} = 0
\lab{sub1tensor}
\ee
and the constraints 
\be
A_{\mu}^{S,r} {\tilde B}^{\mu}_{i}\;  
\sbr{\(\sum_{m=0}^{n-1} 
\(\otimes 1\)^m \otimes S_r \(\otimes 1\)^{n-m-1}\)}{\(\sum_{l=0}^{n-1}
c_{n,l}\;  \(\otimes  P_{\Lambda}\)^l \otimes P_j 
\(\otimes  P_{\Lambda}\)^{n-l-1}\)}  = 0  
\lab{sub2tensor}
\ee

Therefore, since \rf{sub1tensor} are the same equations as \rf{zcmodel}  we 
have a submodel, and  the subclass of
solutions is determined by the constraints \rf{sub2tensor}. 

The conserved currents obtained from the zero curvature are (see
\rf{currents}) 
\br
J^{\l (n)}_{\mu} &\equiv& \( \otimes W^{-1}\)^{n} \, {\tilde B}_{\mu}^{(n)} \, 
\( \otimes W\)^{n} 
\er

Therefore, if we can choose the constants $c_{n,l}$ in such way that
\rf{sub2tensor} imposes the same set of constraints  for any
$n$, then one gets that the corresponding submodel possesses an infinity
number of conserved currents.

\sect{Applications and Outlook}
\label{sect:Out}

The first applications of our zero curvature generalization have been to
produce new geometrical formulations 
of well known theories in higher dimensions.
The most obvious way to have flatness locally is by requiring  each
component of 
$B_{\mu\nu}$ to be covariantly constant and to commute at the same point. This
simplest possibility, which has not been reviewd here,  produces topological 
 theories like Chern-Simons \ct{Labas} . Since these theories are
rather well understood in the quantum case, a possible application is to
compare 
to the classical solution.  Self dual Yang-Mills or BPS are also easily
incorporated and were also given  in \ct{noso} as 4d examples. 

The case which has  been  studied in more detail is when the local equation
are based on a non-semisimple Lie algebra, which has produced many new
results, specially 
obtaining reductions with infinite number of conserved currents.
First it was applied to $CP^1$  in  \ct{noso},   and to the 
chiral ($su(2)$) model in \ct{mad}, later generalized to Grassmannian 
models in \ct{fuji} and then to  homogeneous spaces in general in \ct{fl},
including symmetric spaces (compact and  noncompact). In this last general
formulation, new models and many classes of the previous cases where
treated with new results and insights. In particular, the reason behind the
infinity of  
conserved currents, discussed only  heuristically
 in \ct{noso}, has been fully understood in \ct{fl} with  the coset
construction $G/K$ in terms of the  
branching of representations of $G$ in those of $K$ with some singlets states
playing a special and important role. This has
been  
explained in detail here (see section \ref{sect:intsub}) directly in terms of
the decomposition of the algebra of $G$ in a subalgebra  
$K$ and  its complement. 

One of the most relevant aspects of our approach is the reduction to submodels 
possessing infinite number of conserved charges. Although that may resemble
the BPS condition, it apparently does not involve  saturation of any bound. 
Such methods have also been applied to Skyrme and Skyrme-type models  
\ct{afz,mad2}. The existence of solutions for the constraints is highly non 
trivial, specially for the  hedgehog for Skyrme \ct{mad2}. 
Indeed it is  a non trivial fact that
there are solutions to the constraints, like  the special one found in
\ct{afz} with the rational map, and the  meromorphic
ones for adjusted babyskyrme with an old trick due to Smirnov and Sobolev in
\ct{fuji2}. 

Although there is a possibility to implement  dressing, as mentioned in 
\ct{noso}, the  systematic method to construct solutions, which come naturally
in $2d$ from the affine algebraic structures,   has still to be developed. 
\vspace{1 cm}

\section*{Acknowledgments}
JSG would like to thank the La Plata colleagues for the perfect organization
and all the participants for the friendly and rewarding atmosphere. This work is
partially supported by grants from from CNPq (Brazil), NSF (PHY-9507829), EC 
(TMRERBFMRXCT960012) and DGICYT (PB 96-0960).

\end{document}